# THE SPIN–PERIOD HISTORY OF INTERMEDIATE POLARS

Joseph Patterson,[1,2] Enrique de Miguel,[3,4] Jonathan Kemp,[5,2] Shawn Dvorak,[6] Berto Monard,[7] Franz-Josef Hambsch,[8] Tonny Vanmunster,[9,10] David R. Skillman,[11] David Cejudo,[12] Tut Campbell,[13] George Roberts,[14] Jim Jones,[15] Lewis M. Cook,[16] Greg Bolt,[17] Robert Rea,[18] Joseph Ulowetz,[19] Thomas Krajci,[20] Kenneth Menzies,[21] Simon Lowther,[22] William Goff,[23] William Stein,[24] Matt A. Wood,[25] Gordon Myers,[26] Geoffrey Stone[27], Helena Uthas[28], Emir Karamehmetoglu[29], Jim Seargeant[30], & Jennie McCormick[31]

[1] Department of Astronomy, Columbia University, 550 West 120th Street, New York, NY 10027; jop@astro.columbia.edu
[2] Visiting Astronomer, Cerro Tololo Inter-American Observatory, National Optical Astronomy Observatories, which is operated by the Association of Universities for Research in Astronomy, Inc., (AURA) under cooperative agreement with the National Science Foundation
[3] Departamento de Física Aplicada, Facultad de Ciencias Experimentales, Universidad de Huelva; 0000-0002-1381-8843
[4] CBA–Huelva, Observatorio del CIECEM, Parque Dunar, Matalascañas, 21760 Almonte, Huelva, Spain; edmiguel63@gmail.com
[5] Mittelman Observatory, Middlebury College, Middlebury, VT 05753; jkemp@middlebury.edu; 0000-0002-8675-8079
[6] CBA–Orlando, Rolling Hills Observatory, 1643 Nightfall Drive, Clermont, FL; sdvorak@rollinghillsobs.org
[7] CBA–Kleinkaroo, Klein Karoo Observatory, PO Box 281, Calitzdorp 6660, South Africa; astroberto13m@gmail.com
[8] CBA–Mol, ROAD Observatory, Oude Bleken 12, B-2400 Mol, Belgium; hambsch@telenet.be
[9] CBA–Belgium, Walhostraat 1A, B-3401 Landen, Belgium; tonny.vanmunster@gmail.com
[10] CBA–Extremadura, e-EyE Astronomical Complex, ES-06340 Fregenal de la Sierra, Spain
[11] CBA–East, 159 Research Road, Greenbelt, MD 20770; rick70720@gmail.com
[12] CBA–Madrid, Observatorio El Gallinero, El Berrueco, 28192 Madrid, Spain; davcejudo@gmail.com
[13] CBA–Arkansas, 7021 Whispering Pine, Harrison, AR 72601; jmontecamp@yahoo.com
[14] CBA–Tennessee, 2007 Cedarmont Drive, Franklin, TN 37067; georgeroberts0804@att.net
[15] CBA–Oregon, Jack Jones Observatory, 22665 Bents Road NE, Aurora, OR 97002; nt7t@centurytel.net
[16] CBA–Concord, 1730 Helix Court, Concord, CA 94518; lew.cook@gmail.com
[17] CBA–Perth, 295 Camberwarra Drive, Craigie, Western Australia 6025, Australia; gbolt@iinet.net.au
[18] CBA–Nelson, Regent Lane Observatory, 8 Regent Lane, Richmond, Nelson 7020, New Zealand; reamarsh@slingshot.co.nz
[19] CBA–Illinois, Northbrook Meadow Observatory, 855 Fair Lane, Northbrook, IL 60062; joe700a@gmail.com
[20] CBA–New Mexico, PO Box 1351 Cloudcroft, NM 88317; tom_krajci@tularosa.net
[21] CBA–Framingham, 318A Potter Road, Framingham, MA 01701; kenmenstar@gmail.com
[22] CBA–Pukekohe, Jim Lowther Observatory, 19 Cape Vista Crescent, Pukekohe 2120, New Zealand; simon@jlobservatory.com
[23] CBA–Sutter Creek, 13508 Monitor Lane, Sutter Creek, CA 95685; b-goff@sbcglobal.net
[24] CBA–Las Cruces, 6025 Calle Paraiso, Las Cruces, NM 88012; starman@tbelc.org
[25] Department of Physics & Astronomy, Texas A&M University–Commerce, Commerce, TX 75429; matt.wood@tamuc.edu
[26] CBA–San Mateo, 5 Inverness Way, Hillsborough, CA 94010; gordonmyers@hotmail.com; 0000-0002-9810-0506
[27] CBA–Sierras, 44325 Alder Heights Road, Auberry, CA 93602; geofstone@earthlink.net
[28] Viktor Rydberg–Djursholm, Viktor Rydbergs väg 31, 182 62 Djursholm, Sweden; helena.uthas@vrg.se
[29] Department of Physics and Astronomy, Aarhus University, Ny Munkegade 120, DK-8000 Aarhus C, Denmark; emir.k@phys.au.dk
[30] CBA–Edgewood, 11 Hilltop Road, Edgewood, NM 87015; jimsarge@gmail.com
[31] CBA–Pakuranga, Farm Cove Observatory, 2/24 Rapallo Place, Farm Cove, Pakuranga, Auckland 2012, New Zealand; farmcoveobs@xtra.co.nz

**ABSTRACT**

We report the detailed history of spin-period changes in five intermediate polars (DQ Herculis, AO Piscium, FO Aquarii, V1223 Sagittarii, and BG Canis Minoris) during the 30–60 years since their original discovery. Most are slowly spinning up, although there are sometimes years-long episodes of spin-down. This is supportive of the idea that the underlying magnetic white dwarfs are near spin equilibrium. In addition to the ~40 stars sharing many properties and defined by their strong, pulsed X-ray emission, there are a few rotating much faster ($P<80$ s), whose membership in the class is still in doubt - and who are overdue for closer study.

**Concepts**

Cataclysmic variable stars (203), Classical novae (251), Close binary stars (254), DQ Herculis stars (407), Interacting binary stars (801), Novae (1127), Stellar accretion (1578), Stellar accretion disks (1579)

**Objects**

V* AO Psc, V* BG CMi, V* DQ Her, V* FO Aqr, V* V1223 Sgr

# 1 INTRODUCTION

Intermediate polars (IPs, also called DQ Her stars after the prototype) are magnetic cataclysmic variables with stable periodic signals in optical and X-ray light, and periods typically in the range 1–30 minutes. These periods come from rotation of the radially accreting, magnetic white dwarf (WD), although some also show "sideband" signals which arise from interaction between the spin and orbital clocks. The first was found long ago, in the remnant of Nova Herculis 1934 (Walker 1956, 1961). X-ray telescopes since 1980 have revealed many more, because most IPs radiate most of their energy in X-rays, as accreting matter plunges radially to the WD surface. Patterson (1994, hereafter P94) reviews these stars, and Table 1 of Norton, Wynn, & Somerscales (2006, hereafter NWS) presents a more complete list of class members. The NASA website created by Koji Mukai[32] contains by far the most up-to-date and useful online list of the ~50 class members and their individual properties. Much of the material in Mukai's (2017) review of X-ray emission in cataclysmic variables is also very pertinent to IPs.

By tracking period changes from year to year, one can in principle measure torques on the rotating WD. This can constrain the accretion rate and the WD magnetic moment (P94; NWS). Several authors have attempted this, based on time-series photometry obtained over a baseline of several years. This has been sufficient to yield a rough estimate for the several stars studied: they change their pulse periods on timescales [= $P/(dP/dt)$] around $10^6$ years.

We have carried out such programs over many years, most recently with the globally-distributed small telescopes of the Center for Backyard Astrophysics (CBA, Patterson et al. 2013). Our baselines are very long, usually from the discovery year to the present. About 30 of the ~50 known class members are in our archives: nearly all IPs brighter than $V$~17. Most are monitored 3–15 times per year, with special care to obtain timings early and late in each observing season (which eliminates errors in counting cycles between consecutive seasons).

Faithful tracking of known IPs appears to be less glamorous than the discovery of new class members. Many of the published studies base their period estimates on just one observing season — sufficient to establish the stability of the fast signal, but not to measure period change. In the course of this work, we have also found that some of the published spin ephemerides spanning more than ~3 years are incorrect. The main reason is cycle-count errors between years, because observers tend to disfavor the poorer observing conditions of early and late season. In addition to frustrating the search for $dP/dt$, lack of a reliable long-term ephemeris, or at least accurate period estimation, also hampers interpretation of data at other wavelengths. But full publication of our results will take years to complete, because of volume (>9000 nights so far), and rapid discovery of new IPs makes our task ever more daunting.

So we here present a summary of the period history of five IPs with the longest baseline of observation. These typify the patterns and time scales found in other class members, but are more clearly defined, because the baseline is longer (at least 35 years).

---

[32] The Intermediate Polars, https://asd.gsfc.nasa.gov/Koji.Mukai/iphome/iphome.html

## 2 INDIVIDUAL STARS

### 2.1 Measurements

In studies of period change, it is standard practice to present "O–C curves", which represent a star's departure from a strict constant-period ephemeris (e.g. Breus et al. 2019, or Kreiner 1971 for a very extensive application to close binaries generally). For *orbita*l period changes, it is essentially mandatory, because the period change is so small as to require the full decades-long baseline to yield a tiny measurable effect. Such studies always present precise timings of individual events (e.g., maximum/minimum light, mid-eclipse) and then fit timings with a constant period or low-order polynomial ($P$, $\dot{P}$, maybe $\ddot{P}$).

This technique flounders when the event being timed is not precisely defined ("maximum light"), or is not observed sufficiently often to establish cycle count with certainty, or when the period changes too fast. Intermediate-polar spin history brings each of these problems into play. In addition, as this study will show, the spin periods of these IPs typically **wander** on timescales of years, for no clear reason. Thus the $\dot{P}$ and $\ddot{P}$ terms of a polynomial fit may be mere accidents of the observing interval. Finally, because our study involves contributions from ~25 different telescopes over ~40 years, the merging of data on a common scale can be a problem.

For these reasons, we present the data not as O–C curves, but as period versus time [*P(t)*]. Most of the periods are measured as running averages over 3-year baselines, which span enough time to give good accuracy. For example, a 20 minute signal will execute ~60000 cycles during 3 years (i.e. an interval of ~2.3 years); if each timing is accurate to 0.07 cycles (a reasonable but conservative estimate), then the period is measured to an accuracy of 0.0013 seconds. With many (>10) such timings, the error shrinks to ~0.0008 seconds. These numbers are typical for most of our data..

To illustrate this point, Figure 1 presents a traditional O–C diagram for the 913 s signal in BG Canis Minoris. The observations span 38 years with no uncertainty in cycle count. Generally, the "sheds water" shape of the curve shows that the period decreases throughout. But the rate of decrease is not constant, and there is no simple mathematical expression to describe it. *P(t)*, as described, is probably a better way to render a long history; compare the O–C curve in Figure 1 with the corresponding *P(t)* representation[33] for BG CMi in Figure 2.

### 2.2 DQ Herculis (Nova Herculis 1934)

DQ Her was regarded as "the nova of the century" for most of the 20th century. Among its several first-ever contributions to cataclysmic-variable science was the discovery of strictly coherent 71-second pulses in the light curve (Walker 1956, 1961). This has been tracked continuously ever since: the most recent study is that of Wood et al. (2005). The *P(t)* history is given in Table 1, and shown in Figure 2. The period has decreased continuously since 1954, with a rate of period decrease declining from 26 to 12 µs/yr.

---

[33] The pulse-period history of the high-luminosity pulsating X-ray sources ("X-ray pulsars") is always represented this way (e.g., the many figures and superb analysis in Bildsten et al. 1997). The reason is that the periods change much too fast for O–C analysis; neutron stars are easy to spin up (or down)! With infrequent observation, the same applies to IPs.

DQ Her is exempt from one of the caveats expressed in §2.1, because the cycle count is known with certainty from 1954 to the present. On the other hand, the timing of the 71 s signal is known to present a large, systematic dependence on orbital phase (Warner et al. 1972, Patterson et al. 1978). So a reliable timing should exclude the eclipse and be averaged over the remainder of the orbit. This is true for most of our data; and since each period estimate is based on at least a dozen (usually >20) nights of observation, this concern should not affect our result.

The accuracy of the DQ Her timings is much greater than that of the other stars considered here. This is because the signal is ~13x faster, because there are no sideband signals to confuse matters, because the signal is accurately sinusoidal, and because the observing season is long. We estimate that each 3-year period estimate is accurate to ~0.003 ms.

## 2.3 AO Piscium (H2252–035)

Following DQ Her, AO Psc was the second linchpin in the discovery of intermediate polars. Griffiths et al. (1980) found the star to be a strong X-ray-emitter in the HEAO A-3 data, Subsequent study showed stable optical signals at 3.6 hours and 859 seconds (Patterson & Price 1981), and an X-ray signal at 805 seconds (White & Marshall 1981). These studies identified 805 s as the white dwarf's spin period, 3.6 hours as the orbital period, and 859 s as the lower (in frequency) orbital sideband of the spin frequency, Both of the fast periods are actually present in the optical photometry, and their beat frequency equals the orbital frequency to within 1 part in $10^5$. This three-period structure came to be the standard for IPs, although one of the fast periods is sometimes missing (within observational limits).

In recent years it has been possible to measure the X-ray spectrum to good precision, and IPs are generally found to be very hard sources, with temperatures exceeding 40 keV. This agrees well with the theory that the main power source is radial accretion from gas accreting along magnetic field lines, high above the magnetic pole. The X-ray frequency $f_X$ is taken to be the true spin frequency. For the well-studied cases, the dominant fast optical signal occurs either at $f_X$ or at the lower orbital sideband $f_X - f_{orb}$, as expected for prograde rotation of the white dwarf (i.e., no certifiably retrograde rotators are known).

The spin-period history of AO Psc's dominant optical (859 s) signal is tabulated and tracked in Table 1 and Figure 2. For the last 40 years, the accreting white dwarf has been spinning up at a rate of ~1.7 ms/yr.

## 2.4 FO Aquarii (H2215–086), V1223 Sagittarii (4U1851–31), and BG Canis Minoris (3A0729+103)

After the discovery of AO Psc, three other hard X-ray sources were quickly found to coincide with stars showing spectra and light curves characteristic of cataclysmic variables, and with similar fast periods found in optical photometry.  Steiner et al. (1981) found a 794 s period in V1223 Sgr, Patterson & Steiner (1983, see also Shafter & Targan 1982) found a 1254 s period in FO Aqr, and McHardy et al (1984) found a 913 s period in BG CMi.  Several papers have tracked period changes since then (V1223 Sgr: Jablonski & Steiner 1987; FO Aqr: Patterson et al. 1998, Littlefield et al. 2018; BG CMi: Patterson & Thomas 1993; see also Norton et al. 1992).  After collecting the published data and adding our own ~400 timings, we find the *P(t)* behavior recorded in Table 1 and shown in Figure 2.

Figure 2 is a representative collection.  Most IPs decrease their periods, as one might expect since they are probably accreting gas from a disk, which has higher specific angular momentum than the WD.  But some show episodes of period increase, which can last as long as 35 years or more (V1223 Sgr).  Since the observed timescales of period change ($P/\dot{P}$ from Figure 2) are typically near $10^6$ years — far shorter than the lifetime of the stars — it is generally thought that the stars are near "spin equilibrium", where they vacillate between episodes of spin-up and spin-down.  That is very likely true, although it has never been proven.  If they do vacillate, it can be on timescales as long as 50 years.

## 3  SPIN-PERIOD CHANGE IN THEORY

The prevailing theory for spin-period change in accreting, magnetic compact stars is that of Ghosh & Lamb (1979).  This is reviewed and applied to IPs by NWS, P94, Lamb & Patterson (1983), and Mukai (2017).  Stars with decreasing periods are deemed to be "slow rotators", with the spin-up matter torque of accreting gas exceeding the spin-down torque of the WD's magnetic field lines entangled in the outer, slowly rotating disk.  But many consecutive years of spin-up will move the star to "fast rotator" status, where the WD's field lines entangle in the outer disk and slow the star down.  Thus is created a spin equilibrium, although modulated by any changes in mass-transfer rate – endemic to all cataclysmic variables.

It should be stressed that the *dP/dt* values (the slopes) in Figure 2 do not, in this theory and any plausible theory, represent actual evolution times — but merely some accidental feature of the current era (possibly mass transfer rate; the WD magnetic moment is also critical, but is assumed constant).

Fast-rotator status tends to inhibit accretion, since it invokes a centrifugal barrier.  Therefore we expect that stars will be fainter during the fast-rotator phase — and thus predict that most known IPs should be in their slow-rotator phase (spinning up).  As apparently observed.  Figure 16 of P94 shows the general idea.

## 4  V1223 SAGITTARII

In Figure 2, V1223 Sgr seems to be an exception.  All the other stars show period decrease, with at most small and short-lived episodes of period increase.  And this appears to be generally true for the other several dozen stars in our program (with more fragmentary data; there may be some interesting exceptions).  Now our general idea is that these stars, supposedly near spin equilibrium, spend comparable times spinning up and down; so why is this a surprise? Because a WD spinning down should be in a state of low luminosity, as its flailing magnetic field lines drag in the slowly rotating disk. This explains why most IPs are spinning up (because the stars are hard to detect in a low-luminosity state). Yet V1223 Sgr appears to be one of the most luminous IPs (Beuermann et al. 2004).

So there are still some mysteries to unpack on this subject. One possibility lies in the star's long-term history of brightness fluctuations, which seem to be much larger than those certified in other IPs. Garnavich & Szkody (1988) report a long-lasting "low state" in the 1940s, and Simon (2017, Figures 1 and 3) shows that this large variability in brightness is quite characteristic of the star.

## 5 SHORT-PERIOD COUSINS?

Among the ~50 known IPs, ~35 have been X-ray-selected. They all have common properties: strong and hard X-rays, high-excitation emission lines, spin periods exceeding 3 minutes, and the presence of sideband signals. The commonalities are sufficiently extensive to warrant lumping into one class: intermediate polars.

But there are a few other cataclysmic variables with stable, very short periods: WZ Sge at 27.87 s; AE Aqr at 33.08 s, V533 Her at 63.63 s, V455 And at 67.62 s; and possibly DQ Her itself at 71.06 s. They each have unique quirks which are the subject of many research papers, and it remains unknown[34] if their underlying physics is predominantly that of the IPs. Their principal disqualifier is the absence of strong, hard, pulsed X-rays. But on the other hand, that may merely be the *result* of fast rotation (see Sec 5.7 of Mukai 2017). Considering that most of these fast periodic signals were discovered in the 1970s — before any of the well-credentialed IPs were found — it seems likely that a well-designed search for more candidates would be fruitful.

## 6 SUMMARY AND A LOOK AHEAD

1. Period-versus-time (Figure 2 and Table 1) is for most purposes the best way to illustrate the period changes — rather than the more traditional O–C diagram, which is hard to interpret when the changes are not monotonic. It should be useful to observers attempting to phase their data, and to theorists trying to understand accretion torques in these and related stars (e.g. X-ray pulsars). We would be happy to reduce our analysis burden by furnishing data on any of the ~30 stars to interested researchers.

2. For 4 of 5 stars reported here, and for most IPs with shorter baselines of observation, spin-up is the general rule. But there are spin-down episodes, and the reason is not yet securely known. An extensive and calibrated record of visual or X-ray brightness, over a baseline of years and supplemented by period data of the type reported here, may reveal that reason.

---

[34] DQ Her itself is generally included in the IP class, on grounds of seniority, period stability, pulsed high-excitation emission lines, and a good excuse for concealing X-rays (a binary inclination close to 90°). The others are usually not included in lists — though possibly because lists are often prepared by X-ray astronomers. Without that energy bias, it's possible that all but WZ Sge would qualify.

3. It should be remembered that IPs have an extra piece of physics (thermonuclear explosions) not available to their accreting neutron-star cousins. This could yield some other sources of spin-up: either an accretion torque specifically associated with the elevated $\dot{M}$ following a nova outburst (which would make IPs temporarily "slow rotators" and thus spin up as their magnetospheres are squashed), or a slow contraction (conserving angular momentum) of a WD heated by a recent outburst. These could be called "Blame It On The Bossa Nova" theories (Gormé et al. 1963). The back of a small envelope does not reveal any enormous problem with the energy or angular momentum requirements. And we note from Figure 2 that the observed *P(t)* in DQ Her, interpreted as exponential decay, appears to suggest a time constant of ~60 years — roughly the age of the postnova.

4. In the present century, almost nothing new has been learned about the fastest IPs. This signifies not the maturity of the subject, but mainly the lack of human attention. The stars themselves seem cooperative: some are very nearby, some shout for attention via classical-nova outbursts, and some have a long history of previous work which has never been well-digested in the context of what is now known about the slower IPs. That might be a fruitful subject area for the 2020s.

# 7 ACKNOWLEDGMENTS

To span 40 years, you have to keep observing for 40 years. In addition to all the small telescopes contributing to this paper, there were also ~500 nights on meter-class telescopes, mainly from Kitt Peak, Cerro Tololo, and Lick Observatory. We owe a great debt to those mountain staffs for all the technical (and medical!) help they provided. On the financial side, grants from the NSF (AST-1615456 and AST-1908582), the Research Corporation, and the Mount Cuba Astronomical Foundation have been a key element in this work. The work also includes observations obtained with the Mittelman Observatories 0.5m telescope at New Mexico Skies in Mayhill, New Mexico. Additionally, we would like to thank the Mittelman Family Foundation for its generous support of and substantial impact upon astronomy at Middlebury College.

**REFERENCES**

Beuermann, K. et al. 2004, A&A, 419, 291.
Bildsten, L. et al. 1997, ApJ, 113, 367.
Breus, V., Andronov, I.L., Dubovsky, P., Petrik, K., & Zola, S. 2019, astro-ph 1912.06183.
Garnavich, P. & Szkody, P. 1988, PASP, 100, 1522.
Ghosh, P. & Lamb, F.K. 1979, ApJ, 234, 296.
Gormé, E. (Vocalist), Weil, C. (Lyricist), and Mann, B. (Composer). 1963. Blame it on the Bossa Nova. [Vinyl record]. New York, NY: Columbia Records. Retrieved from https://www.youtube.com/watch?v=PaRlW-jz1QQ
Griffiths, R.E. et al. 1980, MNRAS, 193, 25p.
Jablonski, F. & Steiner, J.E. 1987, ApJ, 323, 672.
Kreiner, J.L. 1971, AcA, 21, 365.
Lamb, D.Q. & Patterson, J. 1983, ASSL, 101, 229L.
Littlefield, C. et al. 2016, ApJ, 833, L93.
McHardy, I. et al. 1984, MNRAS, 210, 663.
Mukai, K. 2017, PASP, 129, 620.
Norton, A.J. et al. 1992, MNRAS, 258, 697.
Norton, A.J., Wynn, G.A., & Somerscales, R.V. 2004, ApJ, 614, 349 (NWS).
Patterson, J. 1994, PASP, 106, 209 (P94).
Patterson, J., Robinson, E.L. & Nather, R.E. 1978, ApJ, 224, 570.
Patterson, J. & Price, C.M. 1981, ApJ, 243, L83.
Patterson, J. & Steiner, J.E. 1983, ApJ, 264, L61.
Patterson, J. & Thomas, G. 1993, PASP, 105, 59.
Patterson, J. et al. 2013, MNRAS, 434, 1902.
Shafter, A.W. & Targan, D.M. 1982, AJ 67, 655.
Simon, V. 2015, Proc. Sci., in The Golden Age of Cataclysmic Variables – III, Palermo, p. 22.
Steiner, J.E. et al. 1981, ApJ, 249, L21.
Walker, M.F. 1956, ApJ, 123, 68.
Walker, M.F. 1961, ApJ, 134, 171.
Warner, B. et al. 1972, MNRAS, 159, 321.
White, N.E. & Marshall, F. 1981, ApJ, 249, L25.
Wood, M.A. et al. 2005, ApJ, 634, 570.

**FIGURE CAPTIONS**

Figure 1. O–C diagram for maximum light of the 913 s signal in BG CMi, relative to an assumed period of 913.48 s. The general shape indicates a spin-up (period decrease) over the 38 years of observation, but with a rate that is not consistent with a simple polynomial fit.

Figure 2. Period versus time for the 5 IPs with longest duration of observation. The baselines for most points are in the range 500–800 days, and the resultant errors (which represent an estimated phase uncertainty of ~0.07 cycles over those baselines) are about the size of the symbols.

**TABLE 1**

**History of Spin Periods**

| DQ Her | | AO Psc | | FO Aqr | | V1223 Sgr | | BG CMi | |
|---|---|---|---|---|---|---|---|---|---|
| Year | Period (s) (71+) | Year | Period (s) (858+) | Year | Period (s) (1254+) | Year | Period (s) (794+) | Year | Period (s) (913+) |
| 1955.48 | 0.065858 | 1980.5 | 0.6893 | 1982.5 | 0.4487 | 1981.4 | 0.3803 | 1982.7 | 0.5055 |
| 1957.06 | 0.065804 | 1981.5 | 0.6860 | 1983.9 | 0.4474 | 1982.8 | 0.3808 | 1983.6 | 0.5041 |
| 1958.20 | 0.065796 | 1983.5 | 0.6838 | 1985.5 | 0.4530 | 1984.1 | 0.3827 | 1984.4 | 0.5017 |
| 1959.1 | 0.065774 | 1986.0 | 0.675 | 1986.7 | 0.4537 | 1985.7 | 0.3832 | 1985.6 | 0.4978 |
| 1968.07 | 0.065564 | 1988.1 | 0.671 | 1987.7 | 0.4514 | 1988.6 | 0.3863 | 1987.1 | 0.496 |
| 1969.65 | 0.065556 | 1990.2 | 0.6670 | 1989.0 | 0.4522 | 1998.5 | 0.3934 | 1988.6 | 0.492 |
| 1970.2 | 0.065540 | 1993.3 | 0.661 | 1990.4 | 0.4505 | 2001.0 | 0.3959 | 1989.6 | 0.4895 |
| 1971.68 | 0.06550 | 1996.3 | 0.6550 | 1991.4 | 0.4494 | 2004.0 | 0.3983 | 1990.2 | 0.4877 |
| 1973.25 | 0.065456 | 1998.3 | 0.6530 | 1992.2 | 0.4480 | 2005.7 | 0.3986 | 1991.4 | 0.4867 |
| 1975.06 | 0.065435 | 2000.3 | 0.6517 | 1993.4 | 0.4459 | 2007.2 | 0.4023 | 1993.7 | 0.4802 |
| 1976.27 | 0.065413 | 2001.3 | 0.6500 | 1995.4 | 0.4394 | 2008.5 | 0.4018 | 1995.4 | 0.4756 |
| 1977.1 | 0.06540 | 2002.3 | 0.6482 | 1996.2 | 0.4334 | 2010.0 | 0.4029 | 1996.9 | 0.4723 |
| 1978.30 | 0.065374 | 2003.3 | 0.6474 | 1997.2 | 0.4272 | 2011.3 | 0.4043 | 1998.8 | 0.4693 |
| 1980.77 | 0.065342 | 2004.3 | 0.6446 | 1998.2 | 0.4206 | 2012.5 | 0.4045 | 2000.4 | 0.4707 |
| 1982.80 | 0.065313 | 2005.3 | 0.6428 | 1999.4 | 0.4138 | 2013.7 | 0.4052 | 2001.5 | 0.4701 |
| 1985.5 | 0.065268 | 2006.3 | 0.6404 | 2000.2 | 0.4066 | 2014.5 | 0.4054 | 2002.6 | 0.4717 |
| 1989.5 | 0.065203 | 2007.2 | 0.6384 | 2001.2 | 0.3982 | 2016.2 | 0.4066 | 2003.6 | 0.4697 |
| 1992.0 | 0.065170 | 2008.3 | 0.6380 | 2002.2 | 0.3916 | 2017.2 | 0.4069 | 2005.3 | 0.4717 |
| 1993.5 | 0.065158 | 2009.3 | 0.6358 | 2003.2 | 0.3841 | | | 2006.7 | 0.4712 |
| 2000.70 | 0.065050 | 2010.3 | 0.6350 | 2004.2 | 0.3820 | | | 2008.5 | 0.4702 |
| 2001.3 | 0.065037 | 2011.3 | 0.6321 | 2005.2 | 0.3753 | | | 2009.7 | 0.4703 |
| 2002.3 | 0.065028 | 2012.3 | 0.6308 | 2006.2 | 0.3714 | | | 2010.5 | 0.4681 |
| 2007.44 | 0.064972 | 2013.3 | 0.6285 | 2007.2 | 0.3614 | | | 2011.4 | 0.467 |
| 2008.34 | 0.064953 | 2014.3 | 0.6285 | 2008.2 | 0.3560 | | | 2013.7 | 0.4658 |
| 2009.58 | 0.064925 | 2015.3 | 0.6256 | 2012.2 | 0.3396 | | | 2016.3 | 0.4653 |
| 2010.59 | 0.064915 | 2016.3 | 0.6239 | 2014.2 | 0.3311 | | | 2017.4 | 0.4644 |
| 2011.94 | 0.064896 | 2017.3 | 0.6219 | 2015.2 | 0.3324 | | | | |
| 2012.85 | 0.064888 | | | 2016.2 | 0.3360 | | | | |
| 2013.52 | 0.064883 | | | 2017.2 | 0.3379 | | | | |
| 2014.65 | 0.064864 | | | | | | | | |
| 2017.0 | 0.064832 | | | | | | | | |

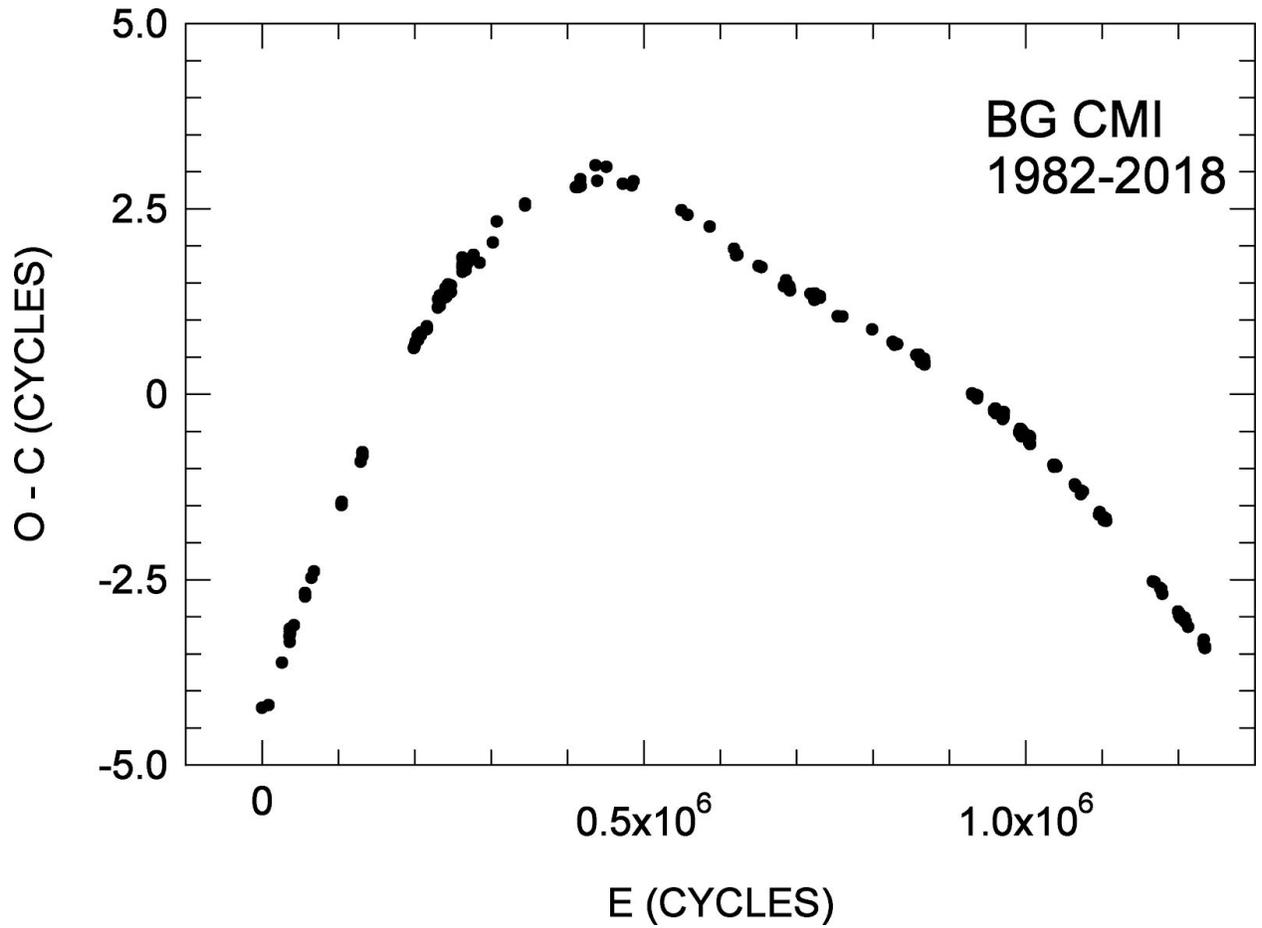

Figure 1. O–C diagram for maximum light of the 913 s signal in BG CMi, relative to an assumed period of 913.48 s. The general shape indicates a spin-up (period decrease) over the 38 years of observation, but with a rate that is not consistent with a simple polynomial fit.

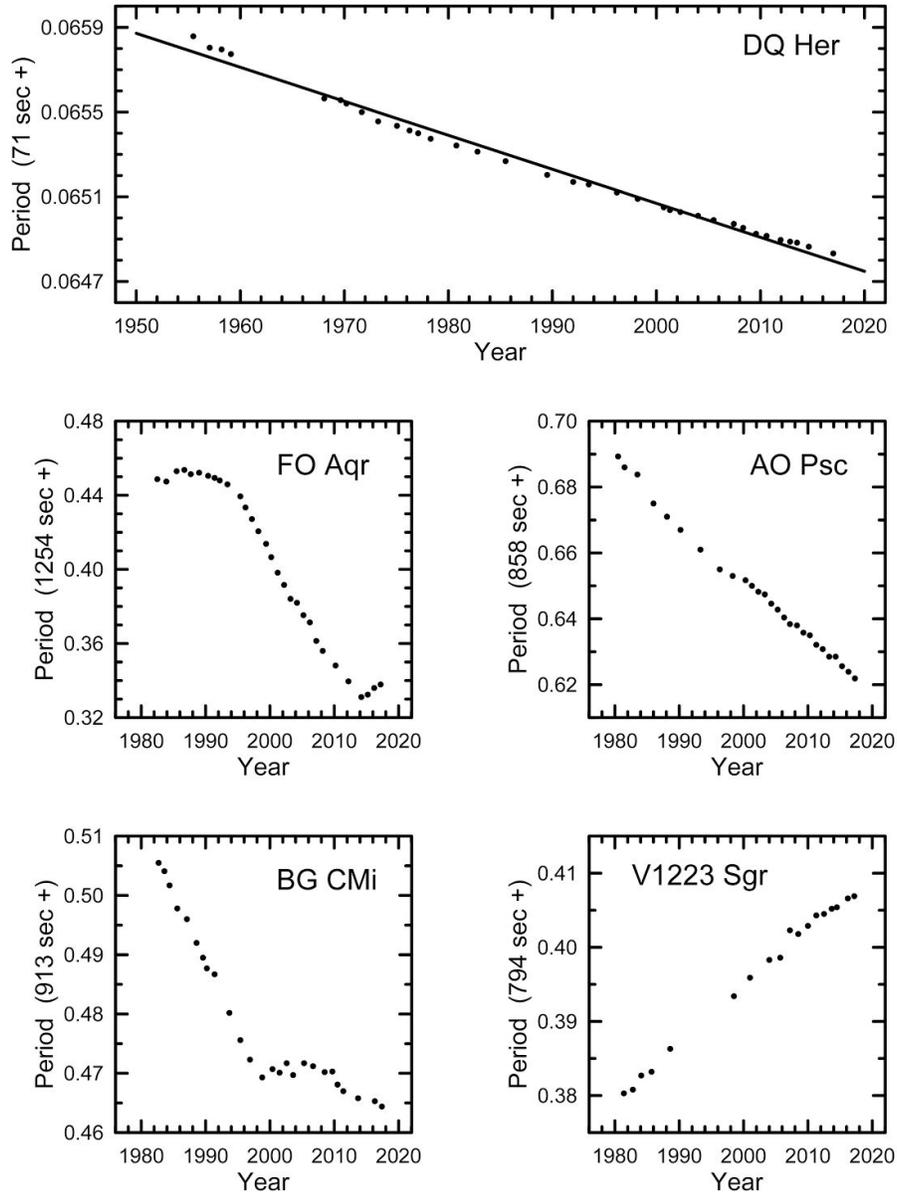

Figure 2. Period versus time for the 5 IPs with longest duration of observation. The baselines for most points are in the range 500–800 days, and the resultant errors (which represent an estimated phase uncertainty of ~0.07 cycles over those baselines) are about the size of the symbols.

13